\documentclass[twocolumn,pra]{revtex4}
\usepackage{graphicx}
\usepackage{epsfig}
\usepackage{latexsym}
\usepackage{amssymb}
\usepackage{bm}
\usepackage{float}
\usepackage{subfigure}
\usepackage{amsmath}
\usepackage{amsfonts}
\begin{document}

\title{The particle in a box in ${\cal PT}$ quantum mechanics and an electromagnetic analog}
\author{Anirudh Dasarathy$^{(1)}$}
\author{Joshua P. Isaacson$^{(2)}$}
\author{Katherine Jones-Smith$^{(3)}$}
\author{Jason Tabachnik$^{(4)}$ }
\author{Harsh Mathur$^{(4)}$}
\affiliation{\vspace{1mm}}
\affiliation{$^{(1)}$ Hawken School, 12465 County Line Road, P.O. Box 8002, Gates Mills, Ohio 44040}
\affiliation{$^{(2)}$ Department of Physics and Astronomy, Michigan State University, 567 Wilson Road, 
East Lansing, MI 48824}
\affiliation{$^{(3)}$ Physics Department and McDonnell Center for the Space Sciences, 
Washington University, St Louis, MO 63130, USA}
\affiliation{$^{(4)}$Department of Physics, Case Western Reserve University, 10900 Euclid Avenue, Cleveland OH 44106}

\begin{abstract}
In ${\cal PT}$ quantum mechanics a fundamental principle of quantum mechanics, that the Hamiltonian must be
hermitian, is replaced by another set of requirements, including notably symmetry under ${\cal PT}$, where 
${\cal P}$ denotes parity and ${\cal T}$ denotes time reversal.
Here we study the role of boundary conditions in ${\cal PT}$ quantum mechanics by
constructing a simple model that is the ${\cal PT}$ symmetric
analog of a particle
in a box. The model has the usual particle in a box Hamiltonian but
boundary conditions that respect ${\cal PT}$ symmetry rather than
hermiticity. We find that for a broad class of ${\cal PT}$ symmetric boundary conditions
the model respects the condition of unbroken ${\cal PT}$ symmetry, namely that
the Hamiltonian and the symmetry operator ${\cal PT}$ have simultaneous
eigenfunctions, implying that the energy eigenvalues are real. We also find that
the Hamiltonian is self-adjoint under the ${\cal PT}$ inner product. Thus
we obtain a simple soluble model that fulfils all the requirements
of ${\cal PT}$ quantum mechanics. In the second part of this paper we formulate a
variational principle for ${\cal PT}$ quantum mechanics that is the analog of  
the textbook Rayleigh-Ritz principle. Finally we consider electromagnetic analogs of the
${\cal PT}$ symmetric particle in a box. We show that the isolated particle in a box may be
realized as a Fabry-Perot cavity between an absorbing medium and its conjugate gain medium.
Coupling the cavity to an external continuum of incoming and outgoing states turns the energy
levels of the box into sharp resonances. Remarkably we find that the resonances have a Breit-Wigner
lineshape in transmission and a Fano lineshape in reflection; by contrast in the corresponding hermitian case the
lineshapes always have a Breit-Wigner form in both transmission and reflection. 

\end{abstract}

\maketitle

\section{Introduction}
\label{sec:introduction}

In ${\cal PT}$ quantum mechanics the canonical principle that the Hamiltonian must be hermitian is relaxed. 
Nonetheless it remains possible to consistently formulate quantum mechanics if the Hamiltonian operator
respects certain conditions, notably ${\cal PT}$ symmetry.
In principle ${\cal PT}$ quantum mechanics expands the
set of Hamiltonians that can be used to describe natural phenomena. The purpose of this paper is
to add to known examples of ${\cal PT}$ quantum mechanics by formulating an especially simple model:
the particle in a box. 

Prior work on ${\cal PT}$ quantum mechanics has considered Hamiltonians 
in which the potential is imaginary \cite{bender, bendereview, benderbox} and hence the Hamiltonian
is manifestly non-hermitian. 
Here we keep the usual particle in a box Hamiltonian.
Non-hermiticity enters the problem through the boundary conditions. 
There is a well developed theory of self-adjoint extensions that determines the boundary conditions that are 
permissible in quantum mechanics if one imposes hermiticity on the Hamiltonian \cite{vonneuman, stone}. 
Here we develop the ${\cal PT}$ counterpart of
this body of knowledge by consideration of a simple example.
In complementary prior work Nelson and Hatano \cite{nelson} have studied a model with non-Hermitian periodic 
boundary conditions. 


Another new result in this paper is the formulation of a variational principle 
that is the ${\cal PT}$ quantum mechanics analog  of the textbook Rayleigh-Ritz variatonal princple.

In a development parallel to, but separate 
from, ${\cal PT}$ quantum mechanics it has been found that photonic structures with
${\cal PT}$ symmetry have many remarkable properties \cite{ptexpt, makris, longhi}, notably the coexistence of lasing and perfect coherent
absorption \cite{cpa1, cpa2, longhi2, chong}. In this paper we also construct a classical electromagnetic analog of the ${\cal PT}$ symmetric particle
in a box. By allowing the modes of the box to couple to an external continuum of incoming and outgoing modes we
obtain an electromagnetic structure that has sharp resonances in place of the bound states of an isolated
box. Remarkably we find that these resonances have a Breit-Wigner lineshape in transmission and a Fano
lineshape in reflection.

\section{Particle in a Box}
\label{sec:box}

\subsection{Boundary conditions}
\label{sec:boundary}

We consider a non-relativistic spinless particle in one dimension with position $x$ confined to lie
in a box of size $L$ so that $0 \leq x \leq L$. The particle is governed by the Hamiltonian
\begin{equation}
h = - \frac{\hbar^2}{2m} \frac{ \partial^2 }{\partial x^2};
\label{eq:boxhamiltonian}
\end{equation}
hereafter we shall work in units wherein $\hbar = 1$ and $m=1$. Parity applied to
the wave function has the effect
\begin{equation}
{\cal P} \psi(x) = \psi ( L - x )
\label{eq:parity}
\end{equation}
while time reversal is the anti-linear operator
\begin{equation}
{\cal T} \psi (x) = \psi^\ast (x).
\label{eq:treversal}
\end{equation}
Note that ${\cal P}^2 = 1$ and ${\cal T}^2 = 1$ while $[ {\cal P}, {\cal T} ] = 0$. 
By the linearity of quantum mechanics the wavefunction must obey the boundary
conditions\footnote{More general linear boundary conditions may be written but we
restrict attention to the physically motivated form eq (\ref{eq:bc}) here.}
\begin{eqnarray}
\psi (0) & = & \lambda_1 \psi' (0);
\nonumber \\
\psi (L) & = & \lambda_2 \psi' (L);
\label{eq:bc}
\end{eqnarray}
where $(\lambda_1, \lambda_2)$ are complex numbers. Thus any boundary
condition is fully characterized by the pair $(\lambda_1, \lambda_2)$. 

Thus far we have described features common to both the canonical particle
in a box and its ${\cal PT}$ symmetric counterpart being introduced here. We now turn 
to the differences.
In canonical quantum mechanics one supposes that the inner product of two states is
defined as
\begin{equation}
( \phi, \psi ) = \int_0^L d x \; \phi^\ast (x) \psi (x)
\label{eq:innertextbook}
\end{equation}
and moreover the Hamiltonian is hermitian or self-adjoint with 
respect to this inner product. In order to ensure that the Hamiltonian
is self-adjoint we must pay attention to the boundary conditions that
are imposed on the wave-functions. Integration by parts reveals that
the self-adjointness condition $(\phi, h \psi) = (h \phi, \psi)$ is fulfilled provided
the surface term
\begin{equation}
\left[ \phi^\ast \psi' - \psi^\ast \phi'  \right]_0^L = 0
\label{eq:surfaceterm}
\end{equation}
vanishes. 

If we impose the boundary condition eq (\ref{eq:bc}) on $\psi$ then it is easy
to verify that one must impose the boundary condition
\begin{eqnarray}
\phi (0) & = & \lambda^\ast_1 \phi' (0);
\nonumber \\
\phi (L) & = & \lambda_2^\ast \phi' (L);
\label{eq:brabc}
\end{eqnarray}
on $\phi(x)$ in order to fulfill the 
surface condition eq (\ref{eq:surfaceterm}). 
Note that in general the boundary conditions we must impose on $\phi$ are different from those
we must impose on $\psi$; the boundary conditions on $\phi$ are said to be the adjoint of the
boundary conditions on $\psi$. 

In the special case that $\lambda_1$ and $\lambda_2$ are real, the same boundary condition is imposed
on $\psi$ and $\phi$, and the boundary condition is said
to be self-adjoint. To ensure that the Hamiltonian eq (\ref{eq:boxhamiltonian}) is hermitian we must
impose self-adjoint boundary conditions on the wavefunctions. Textbook treatments of the particle in a
box tend to focus exclusively on the hard wall case $\lambda_1 = \lambda_2 = 0$ but in fact any member
of the two-parameter family of self-adjoint boundary conditions may be used. Which one should be
used in practice depends on the physics of the problem that is being modeled as a particle in a box.
So long as we use self-adjoint boundary conditions we may be sure that the eigenvalues of the Hamiltonian $h$
are real and the eigenfunctions are complete. 

Now let us put aside considerations of hermiticity and ask what kinds of boundary conditions are 
permissible in ${\cal PT}$ quantum mechanics. The criterion we impose is that the boundary
conditions must be ${\cal PT}$ symmetric in the sense that if $\psi(x)$ obeys the boundary conditions
then so does
\begin{equation}
\xi (x) = {\cal PT} \psi(x) = \psi^\ast (L - x).
\label{eq:ptpsi}
\end{equation}
It is only with ${\cal PT}$ symmetric boundary conditions that the Hamiltonian may properly be said to
commute with the operator ${\cal PT}$, one of three necessary conditions for ${\cal PT}$ quantum mechanics, 
according to the criteria enumerated by Jones-Smith and Mathur \cite{kate}. 

To obtain the form of the ${\cal PT}$ symmetric boundary conditions more explicitly note that eq (\ref{eq:ptpsi})
implies $\xi(0) = \psi^\ast(L)$ and $\xi(L) = \psi^\ast(0)$. Also $\xi'(0) = - \psi^{' \ast}(L)$ and $\xi'(L) = - 
\psi^{' \ast} (0)$. If we now impose that $\psi$ obeys eq (\ref{eq:bc}) we find that $\xi$ automatically follows
suit if and only if $\lambda_2 = - \lambda_1^\ast$. Thus ${\cal PT}$ symmetric boundary conditions form
the two parameter family $(\ell_1 + i \ell_2, - \ell_1 + i \ell_2)$ where $\ell_1$ and $\ell_2$ are real whereas
hermitian boundary conditions correspond to $(\lambda_1, \lambda_2)$ where $\lambda_1$ and 
$\lambda_2$ are real. Note that the two families intersect along the lines $\ell_2 = 0$ and $\lambda_1 = - \lambda_2$
respectively. On the other hand the case that $\ell_1 = 0$ is maximally 
non-hermitian\footnote{Note that the family of ${\cal T}$ symmetric boundary conditions 
coincides with the self-adjoint family whereas the family of ${\cal P}$ symmetric boundary conditions
corresponds to $(\lambda, - \lambda)$ where $\lambda$ is complex. Thus the line of overlap
$\ell_2 = 0$ corresponds to boundary conditions that are separately ${\cal P}$ symmetric, as well
as ${\cal T}$ symmetric, as well as symmetric under the combination ${\cal PT}$. The line
$\ell_1 = 0$ has neither ${\cal P}$ symmetry, nor ${\cal T}$ symmetry, but it does have
${\cal PT}$ symmetry.}.

We now briefly consider the ${\cal PT}$ symmetric generalization of periodic 
boundary conditions. We will show that these boundary conditions lead to a model
studied many years ago by Nelson and Hatano \cite{nelson} in connection with the pinning of 
superconducting vortex lines. 
To this end we consider a particle on a ring.
In canonical quantum mechanics the wavefunction	of the particle	must
obey periodic boundary conditions, or, for a ring threaded by flux,
twisted boundary conditions, given by
\begin{equation}
\psi ( \theta + 2 \pi ) = \lambda \psi ( \theta ).
\label{eq:twist}
\end{equation}
Here $\lambda = \exp ( i \phi )$ where $\phi$ is the flux threading
the ring in units of the Aharonov-Bohm flux quantum and $\theta$ is the
angular co-ordinate around the ring. If we relax the condition of hermiticity it is
permissible to consider $\lambda$ to be an arbitrary complex number, not
necessarily of unit modulus. The criterion we now impose is that the boundary condition 
eq (\ref{eq:twist}) must be ${\cal PT}$ symmetric in the sense that  if $\psi(\theta)$
obeys the boundary condition then so does
\begin{equation}
\xi( \theta) = {\cal PT} \psi(\theta) = \psi^\ast (\theta + \pi).
\label{eq:ptring}
\end{equation}
Imposition of this constraint forces the boundary parameter $\lambda$ to be real. 
Whereas in the hermitian case the parameter $\lambda$ must lie on the unit circle in the
complex plane, in the ${\cal PT}$ symmetric case it must lie on the real axis. 
These are precisely the boundary conditions studied by Nelson and Hatano.
For a free particle Nelson and Hatano's boundary conditions do not lead to real eigenvalues,
except for the hermitian cases $\lambda = \pm 1$, but if there is a potential well,
or a disordered potential, real eigenvalues are obtained for some range of 
$\lambda$, followed by a ${\cal PT}$ transition to complex eigenvalues (dubbed a delocalization 
transition by Nelson and Hatano). It is worth noting that one can further
generalize the Nelson and Hatano model to the case that
the wavefunction and its
derivative at $\theta$ and $\theta + 2 \pi$ are related by a $2 \times 2$ transfer
matrix which may be constrained by either hemiticity or by ${\cal PT}$ symmetry. 
However in this paper we will not investigate periodic boundary conditions
further. Instead we focus on the complementary problem of a particle in a box with
${\cal PT}$ symmetric boundary conditions. 

\subsection{Eigenvalues and eigenfunctions}
\label{eq:eigenvalues}

Now let us compute the eigenvalues and eigenfunctions of the particle in a box subject to ${\cal PT}$
symmetric boundary conditions (hereafter called the ${\cal PT}$ symmetric particle in a box). Since we
are now analyzing a non-hermitian eigenvalue problem there is no guarantee that we will obtain
real eigenvalues or that the eigenfunctions we obtain will be complete. Nonetheless we find that
under a broad set of circumstances the spectrum of the ${\cal PT}$ symmetric particle in a box is
entirely real. Furthermore the corresponding eigenfunctions can therefore be chosen to be simultaneous
eigenfunctions of the Hamiltonian and ${\cal PT}$, a condition called ``unbroken ${\cal PT}$'' in the
literature \cite{bendereview}. Unbroken ${\cal PT}$ is the second condition a Hamiltonian must
fulfill according to the criteria of ref \cite{kate}. Thus we arrive at the important conclusion that the
${\cal PT}$ symmetric particle in a box meets the condition of unbroken ${\cal PT}$ for a broad
range of circumstances.

We attempt solutions of the form
\begin{equation}
\psi = A \exp( i k x ) + B \exp ( - i k x )
\label{eq:ansatz}
\end{equation}
where $k$ may be complex. These solutions are eigenfunctions of the Hamiltonian eq (\ref{eq:boxhamiltonian})
with energy
\begin{equation}
E = \frac{1}{2} k^2.
\label{eq:evalue}
\end{equation}
Note that the energy is real only if $k$ is real or pure imaginary.

Imposing ${\cal PT}$ symmetric boundary conditions leads to the quantization condition
\begin{equation}
\exp( i 2 k L ) = \frac{ 1 - i 2 k \ell_1 - k^2 (\ell_1^2 + \ell_2^2) }{ 1 + i 2 k \ell_1 - k^2 (\ell_1^2 + \ell_2^2) }
\label{eq:quantization}
\end{equation}
and the amplitude ratio
\begin{equation}
\frac{A}{B} = - \frac{ 1 - k \ell_2 + i k \ell_1}{1 + k \ell_2 - i k \ell_1}.
\label{eq:ratio}
\end{equation}
The quantization condition eq (\ref{eq:quantization}) determines the allowed values of $k$ and hence the
energy levels; eq (\ref{eq:ansatz}) and eq (\ref{eq:ratio}) then determine the corresponding eigenfunctions.

Note that in the hermitian case $\ell_2 = 0$ the right hand side of the quantization condition 
has magnitude unity leading to the expected result that the allowed $k$ values and hence the energy levels
are both real. Remarkably for the maximally non-hermitian case $\ell_1 = 0$, the right hand side of the
quantization condition is exactly unity, once again leading to real $k$ values and energy levels.
It is also clear that the only solutions to eq (\ref{eq:quantization}) lie on the real axis in the complex $k$ plane
so long as $\ell_1 > 0$. For $\ell_1 < 0$ it is possible to obtain complex solutions corresponding to broken
${\cal PT}$ symmetry. 

Here for simplicity let us focus on the maximally non-hermitian case $\ell_1 = 0$ where it is easy to see that the 
allowed wave-vectors are 
\begin{equation}
k_n = \frac{\pi n}{L} \hspace{3mm} {\rm with} \hspace{3mm} n = 1, 2, 3, \ldots
\label{eq:kallowedptbox}
\end{equation}
with corresponding energy levels
\begin{equation}
E_n = \frac{ \pi^2 n^2}{2 L^2}.
\label{eq:levelsptbox}
\end{equation}
The corresponding eigenfunctions are
\begin{equation}
\psi_n (x) = N_n \left[ \sin (k_n x) + i k_n \ell_2 \cos (k_n x) \right]
\label{eq:efnsptbox}
\end{equation}
where $N_n$ is a normalization factor to be determined. Thus we see that the energy levels of the
maximally non-hermitian box are identical to the energy levels of the textbook particle in a box with
hard wall boundary conditions. The eigenfunctions however are quite different. 
The relationship between the maximally non-hermitian model and the textbook particle in a box is
discussed further in the appendix. It is easy to verify
that the eigenfunctions eq (\ref{eq:efnsptbox}), are also eigenfunctions of the symmetry operator ${\cal PT}$ with eigenvalue
$(-1)^{n + 1}$. Thus ${\cal PT}$ is unbroken. 

We turn now to the normalization of the eigenfunctions eq (\ref{eq:efnsptbox}). Note that eigenfunctions
corresponding to distinct eigenvalues are not orthogonal with respect to the canonical inner product
eq (\ref{eq:innertextbook}), reflecting the non-hermiticity of the problem. Thus we cannot 
use this inner product for normalization. Although the eigenfunctions of a non-hermitian operator 
are not orthogonal, a weaker result called biorthogonality remains applicable \cite{morse}.
One can show that the eigenvalues of $h$ and its adjoint $h^\dagger$ are complex conjugates of
each other. Moreover if $\psi$ is an eigenfunction of $h$ and $\varphi$ is an eigenfunction of
$h^\dagger$ with eigenvalues that are not complex conjugates of each other, then $\varphi$ and
$\psi$ will be orthogonal. Translated to the present context, the adjoint of a particle in a box with
boundary conditions $(i \ell_2, i \ell_2)$ is a particle in a box with the adjoint boundary conditions
$( - i \ell_2, - i \ell_2 )$. The eigenfunctions of the former problem are given
by eq (\ref{eq:efnsptbox}); the eigenfunctions of the latter by
\begin{equation}
\varphi_n (x) = \tilde{N}_n \left[ \sin ( k_n x ) - i  k_n \ell_2 \cos ( k_n x ) \right].
\label{eq:adjointboxefns}
\end{equation}
These eigenfunctions will respect the biorthonormality
\begin{equation}
( \varphi_n, \psi_m ) = \int_0^L d x \; \varphi_n^\ast (x) \psi_m (x) = \delta_{n,m}
\label{eq:biorthogonality}
\end{equation}
if we choose the normalization factors 
\begin{equation}
\tilde{N}_n N_n = \frac{2}{L} \frac{1}{1 - k_n^2 \ell_2^2 }.
\label{eq:normalization}
\end{equation}
A symmetric way to partition eq (\ref{eq:normalization}) is to choose
\begin{equation}
N_n = \sqrt{ \frac{2}{L} } \frac{1}{ | 1 - k_n^2 \ell_2^2 |^{1/2} }
\label{eq:symmchoice}
\end{equation}
and $\tilde{N}_n = - {{\rm sgn}}(n) N_n$ where the ${{\rm sgn}} (n)$ is the sign of
$k_n^2 \ell_2^2 - 1$.

For some non-hermitian operators the eigenfunctions also satisfy a
bi-completeness relation which in the present context would state
\begin{equation}
\sum_{n=1}^{\infty} \varphi^\ast_n (x) \psi_n (x') = \delta (x - x')
\label{eq:bicompleteness}
\end{equation}
However, bi-completeness is not guaranteed and must be proved on a
case by case basis. We do not at present have a proof that it holds
in the case under consideration here.

In summary, in this subsection we have shown that the ${\cal PT}$ symmetric
particle in a box has unbroken ${\cal PT}$ symmetry for all boundary 
conditions wherein $\ell_1 > 0$. We have explicitly computed the eigenvalues
and eigenfunctions for the maximally non-hermitian case $\ell_1 = 0$
and found that the eigenvalues are real for all $\ell_2$ and that the
eigenfunctions of the Hamiltonian may also be chosen to be eigenfunctions
of the operator ${\cal PT}$. Finally, we have used biorthogonality to normalize
the eigenfunctions. 

\subsection{Inner products}
\label{eq:innerproducts}

In this sub-section we introduce the ${\cal PT}$ inner product and demonstrate
that the Hamiltonian eq (\ref{eq:boxhamiltonian}) subject to ${\cal PT}$ symmetric
boundary conditions is self-adjoint with respect to the ${\cal PT}$ inner product. 
Self-adjointness under the ${\cal PT}$ inner product is the third key condition that
a ${\cal PT}$ quantum Hamiltonian must meet according to ref \cite{kate}. It follows
from ${\cal PT}$ self-adjointness that the eigenfunctions of the ${\cal PT}$ symmetric
particle in a box are orthogonal under the ${\cal PT}$ inner product. At least conceptually,
this eigenfunction orthogonality is different from the biorthogonality discussed above and it
provides another way to normalize the eigenfunctions. These results then allow us to 
define the ${\cal C}$ operator and the associated ${\cal CPT}$ inner product and thereby
complete the formulation of the particle in a box in ${\cal PT}$ quantum
mechanics.

For a finite dimensional system the wavefunction may be represented as a column of complex
numbers denoted $\psi$ and the canonical inner product eq (\ref{eq:innertextbook}) of two states
$\psi$ and $\phi$ may be written as $(\phi, \psi) = \phi^\dagger \psi$. On the other hand the 
${\cal PT}$ inner product is given by
$( \phi, \psi )_{{\cal PT}} = ( {\cal PT} \phi )^T \psi$ \cite{bendereview, kate}. 
Generalizing to a particle in a box we may write the ${\cal PT}$ inner product as
\begin{equation}
( \phi, \psi )_{{\cal PT}} = \int_0^L d x \; \phi^\ast (L - x) \psi (x)
\label{eq:ptinnerprod}
\end{equation}
Eq (\ref{eq:ptinnerprod}) should be contrasted with the standard inner product
eq (\ref{eq:innertextbook}). The ${\cal PT}$ inner product suffers from the defect that it is not
positive definite; nonetheless, it plays an important role in the formulation of ${\cal PT}$
quantum mechanics.

Having defined the ${\cal PT}$ inner product let us now show that the Hamiltonian
eq (\ref{eq:boxhamiltonian}) is self-adjoint under this inner product. To this end we
equate $( \phi, h \psi)_{{\cal PT}} = ( h \phi, \psi )_{{\cal PT}}$. Simple integration by 
parts then reveals this equality will hold provided the surface term
\begin{equation}
\left[ \phi^\ast (L - x) \psi'(x) + \phi'^\ast (L - x) \psi (x) \right]_0^L
\label{eq:ptsurface}
\end{equation}
vanishes.  If we now impose that $\psi$ obeys the ${\cal PT}$ symmetric boundary
condition $( \ell_1 + i \ell_2, - \ell_1 + i \ell_2 )$ we find that in order for the surface
term to vanish, $\phi$ must obey the same boundary condition. Thus we have shown
that the Hamiltonian eq (\ref{eq:boxhamiltonian}) subject to ${\cal PT}$ symmetric
boundary conditions is self-adjoint under the ${\cal PT}$ inner product as claimed. 

An immediate consequence of ${\cal PT}$ self-adjointness is that the eigenfunctions
of the ${\cal PT}$ symmetric particle in a box with distinct eigenvalues should be 
orthogonal under the ${\cal PT}$ inner product. For the maximally non-hermitian
case, $\ell_1 = 0$, a short calculation reveals
\begin{equation}
(\psi_n, \psi_m)_{{\cal PT}} = (-1)^{n+1} \frac{L}{2} (1 - k_n^2 \ell_2^2 ) N_n^2 \delta_{n,m}.
\label{eq:ptorthonormal}
\end{equation}
Here we have made use of eqs (\ref{eq:efnsptbox}) and eq (\ref{eq:ptinnerprod})
but not yet committed ourselves to the choice eq (\ref{eq:symmchoice}) for the
normalization factor. Note that there is a sequence of $\ell_2$ values $\ell_2 = L / \pi n$
where $n = 1, 2, 3, \ldots$ for which we obtain an orthogonality catastrophe: the 
$n^{{\rm th}}$ eigenfunction is self-orthogonal under the ${\cal PT}$ inner product. 
Except at these isolated points we may adopt the normalization eq (\ref{eq:symmchoice})
in which case the expression for the ${\cal PT}$ inner product simplifies to
\begin{equation}
(\psi_n, \psi_m)_{{\cal PT}} = (-1)^n {\rm sgn} (n) \delta_{n.m}
\label{eq:ptorthoclean}
\end{equation}
where ${{\rm sgn}} (n)$ is the sign of $k_n^2 \ell_2^2 - 1$.

Armed with these results we may now define the linear ${\cal C}$ operator via 
\begin{equation}
{\cal C} \psi_n = (-1)^n {{\rm sgn}} (n) \psi_n.
\label{eq:cdefnd}
\end{equation}
Eq (\ref{eq:cdefnd}) fixes the action of the operator ${\cal C}$ in the Hilbert space of states
spanned by the Hamiltonian eigenfunctions $\psi_n$. We shall call this space the physical
space of the ${\cal PT}$ symmetric particle in a box. Within this space the ${\cal C}$ operator
may be written as an integral kernel with the spectral representation
\begin{equation}
{\cal C} (x, x') = \sum_{n = 1}^\infty (-1)^n {{\rm sgn}}(n) \psi_n (x) \varphi_n^\ast (x')
\label{eq:ckernel}
\end{equation}
based on the biorthogonality eq (\ref{eq:biorthogonality}).
Noting that $\varphi_n^\ast(x')$ and $\psi_n (x')$ differ only 
by a factor of $- {\rm sgn}(n)$
we may simplify the expression for ${\cal C}$ to
\begin{equation}
{\cal C} (x, x') = \sum_{n=1}^\infty (-1)^{n+1} \psi_n (x) \psi_n (x')
\label{eq:csimplified}
\end{equation}
where $\psi_n (x)$ is given by eqs (\ref{eq:efnsptbox}) and eq (\ref{eq:symmchoice}).

Finally, following \cite{bendereview, kate}, we define the ${\cal CPT}$ inner product
of two states as
\begin{eqnarray}
( \phi, \psi )_{{\cal CPT}} & = & \int_0^L d x \; ( {\cal CPT} \phi ) (x) \psi (x) 
\nonumber \\
& = &  \int_0^L d x \; \int_0^L d x' \; \psi (x) {\cal C}(x, x') \phi^\ast (L - x').
\nonumber \\
\label{eq:cptinner}
\end{eqnarray}
Evidently for the eigenstates of $h$, by construction, $(\psi_n, \psi_m)_{{\cal CPT}} = \delta_{n,m}$.
Thus the ${\cal CPT}$ inner product is positive definite.

In ${\cal PT}$ quantum mechanics the ${\cal CPT}$ inner product, also known as the dynamically
determined inner product, is deemed the physically correct inner product \cite{bendereview, kate}. 
It is positive definite and time evolution is unitary with respect to it. With the construction of the
${\cal CPT}$ inner product we have therefore completed our formulation of the ${\cal PT}$ symmetric
particle in a box. 

\section{Variational  principle for ${\cal PT}$ Quantum Mechanics}
\label{sec:variational}

The variational principle is a powerful reformulation of Schr\"{o}dinger's equation. It facilitates
the proof of theorems, such as the existence of bound states for arbitrarily weak binding potentials
in one and two dimensions, and it is the basis of fruitful approximation schemes, especially in quantum many-body
physics. It is therefore worthwhile to ask whether there is a similar variational formulation in ${\cal PT}$
quantum mechanics. Remarkably we find that it is possible to reformulate ${\cal PT}$ quantum mechanics
as a variational principle, but only for Hamiltonians that meet the threefold criteria of ${\cal PT}$ 
symmetry, unbroken ${\cal PT}$ and ${\cal PT}$ self-adjointness. 

For comparative purposes it is useful to briefly recall the variational principle in the hermitian case.
For simplicity, consider a finite dimensional Hilbert space in which the state $\psi$ can be
represented as an $N$ component column vector with components $\psi_i$. The Hamiltonian is then
an $N \times N$ matrix with elements $h_{ij}$. The Rayleigh functional $R$ is defined as
\begin{equation}
R = \psi^\dagger h \psi.
\label{eq:rayleigh}
\end{equation}
It is easy to verify that $R$ is real for hermitian $h$. According to the variational principle
the eigenstates of the Hamiltonian are the states that extremise the Rayleigh functional
subject to the normalization constraint $\psi^\dagger \psi = 1$. According to the method
of Lagrange multipliers we must therefore extremise
\begin{equation}
F = \psi^\dagger h \psi - \lambda (\psi^\dagger \psi - 1).
\label{eq:lagrange}
\end{equation}
By setting $\partial F/ \partial \psi_i^\ast = 0$ we recover the Schr\"{o}dinger equation
\begin{equation}h \psi = \lambda \psi.
\label{eq:rayschrodinger}
\end{equation}
On the other hand the condition $\partial F/\partial \psi_i = 0$ 
implies
\begin{equation}
\psi^\dagger h = \lambda \psi^\dagger 
\hspace{3mm}
\Rightarrow
h^\dagger \psi = \lambda^\ast \psi.
\label{eq:conjugaterayschrodinger}
\end{equation}
Note that for $h$ hermitian eqs (\ref{eq:rayschrodinger}) and (\ref{eq:conjugaterayschrodinger}) are
equivalent since $h^\dagger = h$ and $\lambda$ is real. But for $h$ non-hermitian the two equations
are not equivalent and indeed in general are incompatible. Thus hermiticity plays an essential
role in the variational principle. Minimization of $R$ subject to normalization is equivalent to the
Schr\"{o}dinger eigenvalue problem only for hermitian $h$. 

We turn now to the ${\cal PT}$ symmetric case. Again for simplicity we consider a finite 
dimensional Hilbert space. Without loss of generality \cite{kate} we may assume that we are in a basis
such that ${\cal T} \psi = \psi^\ast$ and $P \psi = S \psi$ where $S$ is a diagonal matrix with all its
diagonal entries equal to $\pm 1$. To be definite we assume that our Hilbert space is $2n$ dimensional
and
\begin{equation}
S = \left(
\begin{array}{cc}
{\cal I} & 0 \\
0 & - {\cal I}
\end{array}
\right)
\label{eq:parity}
\end{equation}
where ${\cal I}$ denotes the $n \times n$ identity matrix and $0$ denotes the $n \times n$ matrix with all
entries equal to zero. The ${\cal PT}$ inner product is then given by
\begin{equation}
( \phi, \psi )_{{\cal PT}} = ( {\cal PT} \phi )^T \psi = \phi^\dagger S \psi.
\label{eq:ptinnerfindim}
\end{equation}
The condition of ${\cal PT}$ symmetry, that the Hamiltonian should commute with ${\cal PT}$, implies
$h S = S h^\ast$. This enforces the form
\begin{equation}
h = \left(
\begin{array}{cc}
a & i b \\
i c & d 
\end{array}
\right)
\label{eq:ptblock}
\end{equation}
where $a, b, c$ and $d$ are real $n \times n$ matrices. The condition that the Hamiltonian is self-adjoint
with respect to the ${\cal PT}$ inner product, $(\phi, h \psi)_{{\cal PT}} = ( h \phi, \psi)_{{\cal PT}}$, 
implies $S h = h^\dagger S$, or equivalently $c = b^T$ in eq (\ref{eq:ptblock}).

Armed with this notation we now introduce the ${\cal PT}$ counterpart of the Rayleigh functional 
\begin{equation}
B = ( \psi, h \psi)_{{\cal PT}} = \psi^\dagger S h \psi.
\label{eq:bender}
\end{equation}
To show that the functional $B$ is real it is convenient to write
\begin{equation}
\psi = \left(
\begin{array}{c} 
\xi \\
\eta 
\end{array}
\right)
\label{eq:xieta}
\end{equation}
where $\xi$ and $\eta$ are $n$ component columns.
Then 
\begin{equation}
B = \xi^\dagger a \xi - \eta^\dagger d \eta + i \xi^\dagger b \eta - i \eta^\dagger b^T \xi.
\label{eq:bendercomponents}
\end{equation}
The first two terms are evidently real and the last two are the sum of a conjugate pair,
revealing that $B$ is indeed real.

According to the variational principle for ${\cal PT}$ quantum mechanics we must extremise
$B$ subject to each of three constraints: (a) $(\psi, \psi)_{{\cal PT}} = 1$, (b) $(\psi, \psi)_{{\cal PT}} = 0$
and (c) $(\psi, \psi)_{{\cal PT}} = -1$. To show that the variational principle is equivalent to the 
eigenvalue equation for $h$ let us examine case (a). The other cases can be handled similarly.
By the method of Lagrange multipliers we must look for states $\psi$ that extremise
\begin{equation}
F_B = \psi^\dagger S h \psi - \lambda ( \psi^\dagger S \psi - 1 ).
\label{eq:lagrangebender}
\end{equation}
Imposing $\partial F_B/\partial \psi_i^\ast = 0$ yields the eigenvalue problem 
\begin{equation}
S h \psi = \lambda S \psi \hspace{3mm}
\Rightarrow 
h \psi = \lambda \psi,
\label{eq:bendervalue}
\end{equation}
the desired result, but imposing $\partial F_B /\partial \psi_i = 0$ leads to
\begin{equation}
\psi^\dagger S h = \lambda \psi^\dagger S \hspace{3mm}
\Rightarrow h^\dagger S \psi = \lambda^\ast S \psi.
\label{eq:conjugatebender}
\end{equation}
To show that eq (\ref{eq:conjugatebender}) is equivalent to eq (\ref{eq:bendervalue}),
we use $h^\dagger S = S h$ (${\cal PT}$ self-adjointness)
and the reality of the eigenvalues $\lambda = \lambda^\ast$ (unbroken ${\cal PT}$). 
Thus we see that the three fundamental conditions of ${\cal PT}$ quantum mechanics
play an essential role in the formulation of the variational principle just as hermiticity
does in the hermitian case. 

We conclude with a brief consideration of the variational principle for the particle in a box.
Here we must extremise
\begin{equation}
B = - \frac{1}{2} \int_0^L \; d x \; \psi^\ast (L - x) \frac{ \partial^2 }{ \partial x^2} \psi (x)
\label{eq:benderbox}
\end{equation}
subject to the constraints
$\int_0^L d x \; \psi^\ast (L - x) \psi(x) = -1, 0$ or $1$. 
Making variations with respect to $\psi^\ast$ leads immediately to the 
Schr\"{o}dinger equation
\begin{equation}
- \frac{1}{2} \frac{\partial^2}{\partial x^2} \psi = \lambda \psi.
\label{eq:schrodingervary1}
\end{equation}
Making variations with respect to $\psi$ leads to the equation
\begin{equation}
- \frac{1}{2} \frac{\partial^2}{\partial x^2} \psi^\ast = \lambda \psi^\ast
\label{eq:schrodingervary2}
\end{equation}
provided the surface term
\begin{equation}
\left[ \psi^\ast (L - x) \delta \psi' (x) + \psi'^\ast (L-x) \delta \psi (x) \right]_0^L
\label{eq:varysurface}
\end{equation}
vanishes. The vanishing of the surface term is ensured by requiring the variation $\delta \psi$ 
to obey the same ${\cal PT}$ symmetric boundary conditions as $\psi$. Thus we see the essential
role for the variational principle 
of ${\cal PT}$ symmetric boundary conditions (which ensure both ${\cal PT}$ symmetry and
${\cal PT}$ self-adjointness). Finally we must invoke unbroken ${\cal PT}$  since
eq (\ref{eq:schrodingervary1}) and (\ref{eq:schrodingervary2}) are equivalent only if the
eigenvalues $\lambda$ are real. 

\vspace{3mm}

\section{Electromagnetic Analog}
\label{sec:em}

\subsection{Isolated cavity}
\label{eq:cavity}

In order to construct the electromagnetic analog of the ${\cal PT}$ symmetric particle in a box
let us examine more closely what happens at the boundary points. Close to the left boundary
the wave-function is 
\begin{equation}
\psi = A_1 \exp( i k x) + A_2 \exp( - i k x ).
\label{eq:makingwaves}
\end{equation}
By imposing $\psi (0) = ( \ell_1 + i \ell_2) \psi' (0)$ we conclude that the reflection 
coefficient
\begin{equation}
r_L = \frac{A_1}{A_2} = - \frac{ 1 + k \ell_2 + i k \ell_1}{1 - k \ell_2 - i k \ell_1}.
\label{eq:leftreflect}
\end{equation}
On the other hand if we write eq (\ref{eq:makingwaves}) as 
\begin{equation}
\psi = A_1 \exp ( i k L ) e^{ i k (x - L) } + A_2 \exp( - i k L ) e^{ - i k (x - L)}
\label{eq:shiftwaves}
\end{equation}
and impose the boundary condition $\psi (L) = ( - \ell_1 + i \ell_2 ) \psi' (L)$ we find that the 
right reflection coefficient is 
\begin{equation}
r_R = \frac{A_2 \exp ( - i k L ) }{A_1 \exp( i k L )} = - \frac{1 - k \ell_2 + i k \ell_1}{1 + k \ell_2 - i k \ell_1}.
\label{eq:rightreflect}
\end{equation}
Thus we conclude
\begin{equation}
r_R = \frac{1}{r_L^\ast}.
\label{eq:conjugaterelation}
\end{equation}
The reflection coefficients have the same phase but reciprocal magnitudes.
The quantized energy levels of the box may be determined by solving
\begin{equation}
r_L r_R \exp( i 2 k L ) = 1.
\label{eq:rquantization}
\end{equation}
This is identical to the quantization condition deduced earlier, eq (\ref{eq:quantization}). 
Eq (\ref{eq:conjugaterelation}) is the key feature of the ${\cal PT}$ symmetric particle in a box
that we will seek to emulate in the electromagnetic analog. 

To construct the electromagnetic analog first imagine that the half space $x < 0$ is 
occupied by an absorbing medium while $x > 0$ is void. The electromagnetic field to the right of the
interface may be written as a sum of an incident and reflected wave as
\begin{eqnarray}
E_y & = & {\cal E}_1 e^{ikx} + {\cal E}_2 e^{-ikx},
\nonumber \\
H_z & = & \frac{ {\cal E}_1 }{c \mu_0} e^{i k x} - \frac{ {\cal E}_2 }{c \mu_0} e^{- i k x}.
\label{eq:emrightabsorb}
\end{eqnarray}
We consider only the case of normal incidence and linear polarization here. 
The field to the left of the interface may be written as 
\begin{eqnarray}
E_y & = & \beta \exp^{ - i n k x }, 
\nonumber \\
H_z & = & - \frac{ \beta n}{c} \frac{1}{\mu_0 \mu_r} e^{- i k n x }.
\label{eq:emleftabsorb}
\end{eqnarray}
Here $n = n' + i n'' = \sqrt{\epsilon_r \mu_r}$ is the reflective index of the medium.
We assume $n'' > 0$ corresponding to an absorbing medium. 
Note that the wave eq (\ref{eq:emleftabsorb}) is evanescent and decays inside the
absorbing medium. The amplitude of the wave $\beta$ is determined by the continuity
of $E_x$ and $H_y$ across the interface \footnote{Sometimes in the literature the condition
is imposed that $E_x$ and its first derivative are continuous across the interface. The latter
is equivalent to the continuity of $H_y$ only if the medium in assumed non-magnetic and has $\mu_r = 1$.}.
Performing the match we find that the reflection coefficient is 
\begin{equation}
r_{L {\rm em}} = \frac{ {\cal E}_1 }{ {\cal E}_2 } = \frac{ 1 - n/\mu_r }{1 + n/\mu_r }.
\label{eq:emrl}
\end{equation}

Next imagine that the half space $x > 0$ is occupied by an active medium while $x<0$ is empty.
The active medium is assumed to have refractive index $n^\ast$ and permeability $\mu_r^\ast$
conjugate to the absorbing medium considered in the preceding paragraph. Since $n'' < 0$,
this medium is active. The electromagnetic field to the left of the interface may be written as a 
sum of an incident and reflected wave exactly as in eq (\ref{eq:emrightabsorb}). The field to the right,
inside the medium, is given by
\begin{eqnarray}
E_y & = & \alpha \exp( - i k n^\ast x )
\nonumber \\
H_z & = & - \alpha \frac{n^\ast}{c} \frac{1}{\mu_0 \mu_r^\ast} \exp ( - i k n^\ast x ).
\label{eq:emrightactive}
\end{eqnarray}
Note that the field in the medium is evanescent and decays away from the interface. 
By continuity of $E_x$ and $H_y$ across the interface we may calculate the amplitude
$\alpha$ and the reflection coefficient
\begin{equation}
r_{R \; {\rm em}} = 
\frac{ {\cal E}_2 }{ {\cal E}_1 } =  \frac{ 1 + n^\ast/\mu_r^\ast}{1- n^\ast/\mu_r^\ast}.
\label{eq:rrem}
\end{equation}
Note that $r_{L \; {\rm em}}$ and $r_{R \; {\rm em}}$ have the desired inverse conjugate relationship
eq (\ref{eq:conjugaterelation}). 

It follows that if we build a structure wherein the space $x < 0$ is occupied by the absorbing 
medium, 
the space $x > L$ is occupied by its conjugate active medium, and the slot $0 < x < L$ is left
empty, we will obtain an electromagnetic analog of the ${\cal PT}$ symmetric particle in a box. The electromagnetic
cavity will support undamped oscillations at wave-vectors that meet the quantization condition
eq (\ref{eq:rquantization}), where the reflection coefficients are given by eqs (\ref{eq:emrl})
and (\ref{eq:rrem}). 

\subsection{Scattering Analysis}
\label{sec:scatter}

One way to realize a particle in a box in semiconductor physics is to create a double barrier structure,
for example, a GaAs/AlGaAs sandwich \cite{davies}. In the limit that the barriers are infinitely high a 
particle localized in between is essentially a particle in a box. For finite barrier height
however the eigenstates of the particle in a box broaden into long lived resonances
that can be mapped out by measuring the transmission through the double
barrier structure as a function of energy. The resonances appear as Lorentzian peaks
in the transmission plotted against energy. The natural ${\cal PT}$ symmetric analog of the
double barrier model is a two slab structure in which one slab is absorbing, the other,
its active conjugate. 

In quantum mechanics scattering is powerfully constrained by current conservation which
leads to unitarity for the scattering matrix and pseudo-unitarity [U(1,1) symmetry] for the
transfer matrix. The structure we analyze is instead constrained by ${\cal PT}$ symmetry. 
Moreover it is built of components that do not individually respect ${\cal PT}$ symmetry,
though, for simplicity, we assume that they do respect parity. 
In the remainder of this section we discuss the form of $S$ and $T$ matrices
in the absence of current conservation and with reduced symmetry. In the next subsection these results are used to
analyze resonant scattering from the ${\cal PT}$ symmetric two-slab structure. 

First let us consider a single slab. In general the slab may be illuminated from both sides. 
Thus the field far from the slab is given by
\begin{eqnarray}
E_y & = & A e^{i k x} + B e^{- i k x} \hspace{3mm} {\rm as} \hspace{3mm} x \rightarrow - \infty
\nonumber \\
& = & C e^{i k x} + D e^{- i k x} \hspace{3mm} {\rm as} \hspace{3mm} x \rightarrow + \infty.
\label{eq:fields}
\end{eqnarray}
As before we consider only normal incidence and linearly polarized waves. 
By linearity it follows that the field amplitudes to the left of the slab are related to those 
on the right via the transfer matrix
\begin{equation}
\left(
\begin{array}{c}
C \\
D
\end{array}
\right) = T 
\left( 
\begin{array}{c}
A \\
B
\end{array}
\right).
\label{eq:transferdefined}
\end{equation}
We now make one simplifying assumption, namely, that the slab is parity symmetric. 
Parity applied to the field of eq (\ref{eq:fields}) leads to a field in which the amplitudes
to the left and right are exchanged thus: $A \leftrightarrow D$ and $B \leftrightarrow C$.
Thus symmetry under parity has the consequence that
\begin{equation}
\sigma_x T \sigma_x = T^{-1}.
\label{eq:paritytransfer}
\end{equation}

Eq (\ref{eq:paritytransfer}) implies that $\det T = \pm 1$. The case $\det T = -1$ leads inexorably
to the unphysical result $T = \pm \sigma_x$. Thus we focus on the physical case $\det T = 1$. 
Eq (\ref{eq:paritytransfer}) then leads to the result
\begin{equation}
T = \left( 
\begin{array}{cc}
a & b \\
-b & d 
\end{array} \right).
\label{eq:abbd}
\end{equation}
Here the matrix elements $a, b$ and $ d$ are complex and satisfy $ad + b^2 = 1$. Thus
a transfer matrix that is constrained only by parity has four real parameters. 

The $S$-matrix relates the outgoing amplitudes $(C, B)$ to the incoming amplitudes
$(A, D)$ via
\begin{equation}
\left( 
\begin{array}{c}
C \\
B 
\end{array}
\right) = S
\left(
\begin{array}{c} 
A \\
D
\end{array}
\right).
\label{eq:smatrix}
\end{equation}
We write the elements of the $S$ matrix as
\begin{equation}
S = \left(
\begin{array}{cc}
t_L & r_R \\
r_L & t_R
\end{array}
\right).
\label{eq:selements}
\end{equation}
$t_L$ and $r_L$ are the transmission and reflection amplitudes when there is an incoming wave
from the left $(A = 1, D=0)$; $t_R$ and $r_R$ for the case that the incoming wave is from the right
$(A= 0, D= 1)$. Straightforward algebra reveals that for the parity symmetric transfer matrx
eq (\ref{eq:abbd}) the corresponding $S$ matrix is 
\begin{equation}
S = \frac{1}{d} \left(
\begin{array}{cc} 
1 & b \\
b & 1
\end{array}
\right).
\label{eq:stransfer}
\end{equation}
It is evident from this form that the eigenvalues of the $S$ matrix are 
\begin{equation}
z_1 = \frac{1+b}{d}, \hspace{3mm}
z_2 = \frac{1-b}{d}.
\label{eq:zvalues}
\end{equation}
The associated eigenvectors are respectively the symmetric vector
$\left( \begin{array}{c}1 \\ 1 \end{array} \right) $
and the antisymmetric vector
$\left( \begin{array}{c} 1 \\ -1 \end{array} \right)$. 

\vspace{2mm}

Using eq (\ref{eq:zvalues}) it is straightforward to write the elements of $T$ in terms of $z_1$ and $z_2$ thus
\begin{equation}
a = \frac{2 z_1 z_2}{z_1 + z_2}, \hspace{2mm}
b = \frac{z_1 - z_2}{z_1 + z_2}, \hspace{2mm}
d = \frac{2}{z_1 + z_2}.
\label{eq:abdz}
\end{equation}
The expression for $a$ results from rearranging the constraint
$ad + b^2 = 1$ as $a = (1+b)(1-b)/d = z_1 z_2 d$. 
Thus we may treat $z_1$ and $z_2$ as two basic complex parameters in terms of 
which the transfer matrix may be written. The virtue of using $z_1$ and $z_2$ as the 
basic parameters is that they have a transparent physical interpretation. 
In quantum mechanics the $S$ matrix is unitary and hence its eigenvalues are unimodular.
Hence $|z_1|  = |z_2| = 1$ corresponds to the unitary case. The case that 
$|z_1| < 1$ and $|z_2| < 1$ corresponds to a strictly absorbing slab; the
case that 
$|z_1| > 1$ and $|z_2| > 1$ corresponds to a strictly gainful one. 

In the following it will sometimes be convenient 
to decompose $z_1$ and $z_2$ into their magnitudes and phases as
\begin{equation}
z_1 = \rho e^{\mu} e^{i \phi_1}, \hspace{3mm}
z_2 = \rho e^{-\mu} e^{i \phi_2}.
\label{eq:rlphi}
\end{equation}
We will refer to $\rho$ as the mean absorbance and to $\mu$ as the asymmetry parameter. 

In the quantum or unitary case $\rho =  1$ and $\mu = 0$. 
In that case the expression for the $T$ matrix simplifies to
\begin{eqnarray}
T & = & \left[ 
\begin{array}{cc}
\sec \left( \frac{\phi_1 - \phi_2}{2} \right) e^{\frac{1}{2} ( \phi_1 + \phi_2 )} & i \tan \left( \frac{ \phi_1 - \phi_2 }{2} \right) \\
- i \tan \left( \frac{ \phi_1 - \phi_2 }{2} \right) & \sec \left( \frac{\phi_1 - \phi_2}{2} \right) e^{- \frac{1}{2} ( \phi_1 + \phi_2 )}
\end{array}
\right].
\nonumber \\
\label{eq:unitaryphases}
\end{eqnarray}
We are at liberty to choose the phases $\phi_1$ and $\phi_2$ so that $ - \pi < \phi_1 - \phi_2 \leq \pi$.
Then it is permissible to write 
\begin{eqnarray}
\sec \left( \frac{ \phi_1 - \phi_2 }{2} \right) & = & \cosh \theta, \nonumber \\
\tan \left( \frac{ \phi_1 - \phi_2 }{2} \right) & = & \sinh \theta, \nonumber \\
e^{i \frac{1}{2} ( \phi_1 + \phi_2 ) } & = & e^{i \phi},
\label{eq:quantumpar}
\end{eqnarray}
and thereby replace the parameters $(\phi_1, \phi_2)$ with new parameters 
$(\theta, \phi)$. In terms of these parameters
eq (\ref{eq:unitaryphases}) becomes
\begin{equation}
T = \left( 
\begin{array}{cc}
\cosh \theta e^{i \phi} & i \sinh \theta \\
- i \sinh \theta & \cosh \theta e^{-i \phi} 
\end{array}
\right).
\label{eq:unitaryquantum} 
\end{equation}
The corresponding $S$ matrix is 
\begin{equation}
S = \left( 
\begin{array}{cc}
{\rm sech} \;  \theta e^{i \phi} & i {\rm tanh} \; \theta e^{i \phi} \\
i {\rm tanh} \;  \theta & {\rm sech} \; \theta e^{i \phi}
\end{array}
\right).
\label{eq:unitarysq}
\end{equation}
Comparing to eq (\ref{eq:selements}) we see that the corresponding 
transmission probability is $1/\cosh^2 \theta$. Thus $\theta$ is a measure of the
penetrability of the barrier via quantum tunneling, with large $\theta$ corresponding to
an impenetrable barrier. 

Thus it will prove most transparent to write the parity symmetric transfer matrix eq (\ref{eq:paritytransfer}) 
in terms of the parameters $\rho, \mu, \theta$ and $\phi$. Making use of eqs (\ref{eq:abdz})
and (\ref{eq:rlphi}) and (\ref{eq:quantumpar}) we obtain
\begin{eqnarray}
T & = &
\frac{1}{\cosh \mu + i \sinh \mu \sinh \theta}
\nonumber \\
& \times &
\left[ 
\begin{array}{cc}
\rho \cosh \theta e^{i \phi} & \sinh \mu + i \cosh \mu \sinh \theta \\
- \sinh \mu - i \cosh \mu \sinh \theta & \frac{1}{\rho} \cosh \theta e^{- i \phi} 
\end{array}
\right].
\nonumber \\
\label{eq:muthetarphitransfer}
\end{eqnarray}
Note that eq (\ref{eq:muthetarphitransfer}) reduces to eq (\ref{eq:unitaryquantum}) in the limit
$\rho = 1$ and $\mu = 0$. 

Eq (\ref{eq:muthetarphitransfer}) parametrizes the transfer matrix of any structure that
respects parity. In the appendix we compare the general expression to 
the transfer matrix of a simple rectangular slab of refractive index $n$ and permeability $\mu_r$. 

Having parametrized the transfer matrix of a single slab we now determine $T^R$,  the transfer
matrix of its time reversed counterpart. Time reversing the field configuration eq (\ref{eq:fields})
yields
\begin{eqnarray}
E_y & = & A^\ast e^{-i k x} + B^\ast e^{ i k x} \hspace{3mm} {\rm as} \hspace{3mm} x \rightarrow - \infty
\nonumber \\
& = & C^\ast e^{-i k x} + D^\ast e^{i k x} \hspace{3mm} {\rm as} \hspace{3mm} x \rightarrow + \infty.
\label{eq:reversedfields}
\end{eqnarray}
It follows that
\begin{equation}
\left( 
\begin{array}{c}
D^\ast \\
C^\ast
\end{array}
\right) = T^R \left(
\begin{array}{c}
B^\ast \\
A^\ast
\end{array}
\right)
\label{eq:trdefined}
\end{equation}
where $T^R$ is the transfer matrix of the time reversed slab. 
Comparing eq (\ref{eq:trdefined}) to eq (\ref{eq:transferdefined}) we conclude that
\begin{equation}
T^R = \sigma_x T^\ast \sigma_x, 
\label{eq:trevaluated}
\end{equation}
the result we sought. 

Using eqs (\ref{eq:abbd}), (\ref{eq:abdz}) and (\ref{eq:trevaluated})
it is easy to verify that if $T$ is characterized by the parameters $(z_1, z_2)$,
then the parameters of $T^R$ are $(1/z_1^\ast, 1/z_2^\ast)$. 
Physically this means that if $T$ is absorbing ($|z_1| < 1$ and $|z_2| < 1$) then
$T^R$ has gain ($|z_1| > 1$ and $|z_2| > 1$) and vice-versa, as one might expect intuitively. 
Note furthermore that if we impose time-reversal symmetry, $T = T^R$, we obtain a unitary
$S$ matrix since $z_1$ and $z_2$ are unimodular. 

Explicitly if $T$ is given by eq (\ref{eq:muthetarphitransfer}), then $T^R$ is given
by
\begin{eqnarray}
T^R & =  & \frac{1}{\cosh \mu - i \sinh \mu \sinh \theta} \nonumber \\
& \times  & \left[ 
\begin{array}{cc}
\frac{1}{\rho} \cosh \theta e^{i \phi} & - \sinh \mu + i \cosh \mu \sinh \theta \\
\sinh \mu - i \cosh \mu \sinh \theta & \rho \cosh \theta e^{- i \phi}
\end{array}
\right].
\nonumber \\
\label{eq:trmurhothetaphi}
\end{eqnarray}

Finally we note that if the slab is shifted to the right by a distance $\delta$, its transfer matrix
changes according to 
\begin{equation}
T \rightarrow U(\delta) T U^\dagger (\delta)
\label{eq:shift}
\end{equation}
where the shift matrix 
\begin{equation}
U (\delta) = \left( 
\begin{array}{cc}
e^{- i k \delta} & 0 \\
0 & e^{i k \delta} 
\end{array}
\right).
\label{eq:shiftmatrix}
\end{equation}

\subsection{Transmission spectroscopy of a ${\cal PT}$ symmetric double barrier}
\label{sec:spectroscopy}

We construct a structure with an absorbing slab at $x=-\delta/2$ and its time-reversed
counterpart at $x = \delta/2$. The transfer matrix, $T^D$, of the composite structure is given by
\begin{equation}
T^D = U( - \delta/2 ) T U^\dagger (- \delta/2 ) U ( \delta/2 ) T^R U^\dagger (\delta/2).
\label{eq:td}
\end{equation}
Here $T$ is the transfer matrix of the absorber if it were located at $x=0$; $T^R$, of its 
time reversed counterpart, at the same location. It is easy to verify that $T^D$ satisfies the 
condition of ${\cal PT}$ symmetry $T^D T^{D \ast} = 1$ \cite{longhi2, chong}. In addition
$\det T^D = 1$ since it is a product of matrices with unit determinant. From the transfer matrix
$T^D$ we can easily construct the $S$-matrix $S^D$ which is more directly connected to the
results of scattering experiments. 

It is useful to first consider the unitary case $\rho = 1, \mu = 0$. In this case
both $T$ and $T^R$ are given by eq (\ref{eq:unitaryquantum}). A short calculation reveals
that $t_L = t_R = t$, given by
\begin{equation}
t = \frac{ e^{i 2 \phi} }{ \cosh^2 \theta + \sinh^2 \theta e^{ i 2 k \delta} e^{i 2 \phi} }.
\label{eq:tunitary}
\end{equation}
Eq (\ref{eq:tunitary}) reveals the classic textbook resonance phenomenon. For
\begin{equation}
e^{i 2 k \delta } e^{i 2 \phi} = -1
\label{eq:unitaryresonance}
\end{equation}
we obtain perfect transmission, $|t |^2 = 1$. 
To analyze this resonance write $k = k_c + q$ where $k_c$ is the resonant
wave-vector that meets the condition eq (\ref{eq:unitaryresonance}) and $q$ is the 
detuning. Close to resonance and in the limit of low penetrability (large $\theta$),
the expression for $t$ simplifies to yield the classic Lorentzian lineshape
\begin{equation}
| t |^2 = \frac{1}{ 1 + q^2/Q^2}
\label{eq:lorentzian}
\end{equation}
where $1/Q = 2 \delta \sinh^2 \theta$. In deriving eq (\ref{eq:lorentzian}) we assume
that $(\theta, \phi)$ are independent of $q$, justified a posteriori if the resonance is sufficiently
narrow \footnote{A better but qualitatively similar approximation is obtained by allowing $\phi$ to vary linearly with q.}. Since $|t|^2 + |r|^2 = 1$, the reflection shows a corresponding feature as well. 

Next up in complexity let us suppose the absorbing slab has $\rho < 1$ but the 
asymmetry $\mu =0$. Making the same approximation, in this case the formula for
the $S$-matrix simplifies to 
\begin{equation}
S = \frac{ e^{i 2 \phi} }{1 - i q/Q} \left[
\begin{array}{cc}
1 & \pm i (q/Q) (1/\rho) \\
\pm i \rho (q/Q) & 1 
\end{array}
\right].
\label{eq:sbalance}
\end{equation}
where the $-$ sign applies if the resonant numerator satisfies $\exp ( i k_c \delta + i \phi) = + i$
(even resonance), the $+$ sign if $\exp ( i k_c \delta + i \phi) = - i$ (odd resonance). From the
$S$-matrix we see that there is again a Lorentzian peak in transmission given by eq 
(\ref{eq:lorentzian}). Furthermore the transmission is the same for incidence from the left
or the right; $|t_L|^2 = |t_R|^2 = |t|^2$. The reflection however breaks left-right symmetry
\begin{equation}
|r_L|^2 = \rho^2 \frac{q^2}{Q^2 + q^2}, \hspace{3mm}
|r_R|^2 = \frac{1}{\rho^2} \frac{q^2}{Q^2 + q^2}.
\label{eq:lrreflection}
\end{equation}
Since $\rho < 1$, the reflection for radiation incident from the right is enhanced relative to 
incidence from the left. Note that in the former situation the radiation is incident directly on the
active slab; in the latter on the absorbing slab first. The absorption may be computed from the
conservation law $|t|^2 + |r|^2 + |a|^2 = 1$. Thus we find
\begin{equation}
| a_L |^2 = (1 - \rho^2) \frac{q^2}{Q^2 + q^2}, \hspace{3mm}
| a_R |^2 = \left( 1 - \frac{1}{\rho^2} \right) \frac{q^2}{Q^2 + q^2}.
\label{eq:lrabsorption}
\end{equation}
Note that $|a_R|^2$ is negative, corresponding to a net generation of radiation rather than
absorption, in the case of incidence from the right. 

It is instructive to calculate the eigenvalues and eigenvectors of the $S$-matrix eq (\ref{eq:sbalance}).
Focussing on the case of even resonance we find that the eigenvalues of the $S$-matrix are 
$e^{i 2 \phi}$ and $e^{i 2 \phi} e^{i 2 \xi}$ where $\tan \xi = q/Q$. The corresponding eigenvectors are
\begin{equation}
v_1 = \frac{1}{\sqrt{1 + \rho^2} } \left(
\begin{array}{c}
1 \\
\rho 
\end{array}
\right), 
\hspace{3mm}
v_2 = \frac{1}{\sqrt{1 + \rho^2} } \left(
\begin{array}{c}
1 \\
- \rho 
\end{array}
\right),
\label{eq:v1v2}
\end{equation}
respectively. Remarkably, although the $S$-matrix is not unitary, its eigenvalues are unimodular,
a circumstance dubbed unbroken ${\cal PT}$ by ref \cite{chong}. Note that the eigenvectors
eq (\ref{eq:v1v2}) are not orthogonal in the sense $v^\dagger_1 v_2 = 0$ as they would be 
for a unitary $S$-matrix.  

Unbroken ${\cal PT}$ has the
physical interpretation that absorption happens by interference. If the incident radiation is in either
eigenmode, eq (\ref{eq:v1v2}), it will emerge unattenuated from the structure. 
If however the incident radiation is a superposition $\alpha v_1 + \beta v_2$, then the incident
power is 
\begin{equation}
| \alpha |^2 + | \beta |^2 + \frac{1 - \rho^2}{1 + \rho^2} ( \alpha^\ast \beta + \beta^\ast \alpha )
\label{eq:incidentpower}
\end{equation}
whilst the transmitted power is 
\begin{equation}
| \alpha |^2 + | \beta |^2 + \frac{1 - \rho^2}{1 + \rho^2} ( \alpha^\ast \beta e^{i 2 \xi}  + \beta^\ast \alpha e^{-i 2 \xi} ).
\label{eq:transmittedpower}
\end{equation}
Note that both expressions contain interference terms by virtue of the non-orthogonality of the $S$-matrix 
eigenmodes and this interference is modified by passage through the structure. 

Finally we turn to the most general case $\rho < 1$ and permit non-zero asymmetry $\mu$.
Again we find $t_L = t_R = t$. Right on resonance we find the remarkable effect of 
super-unitary transmission, $| t |^2 \geq 1$. More explicitly, we find $|t|^2 = 1/Z^2$ where 
\begin{equation}
Z = \frac{ 1 - \sinh^2 \mu \cosh^2 \theta }{1 + \sinh^2 \mu \cosh^2 \theta }.
\label{eq:z}
\end{equation}
Near resonance we find 
\begin{equation}
|t|^2 = \frac{1}{ Z^2 + q^2/Q^2},
\label{eq:generalt}
\end{equation}
a Lorentzian with reduced oscillator strength. Here $Q$ is defined as
\begin{equation}
Q = \frac{1}{2 \delta} \frac{ 1 + \sinh^2 \mu \cosh^2 \theta}{ \sinh^2 \theta + \sinh^2 \mu \cosh^2 \theta}.
\label{eq:qdelta}
\end{equation}
Although the transmission resonance remains Lorentzian, the reflection is found to be
\begin{equation}
|r_L|^2  =  \rho^2 \frac{ ( \Delta + q/Q_2 )^2 }{ (1 + q^2/Q^2)^2 }, \hspace{3mm}
| r_R |^2 = \frac{1}{\rho^2} \frac{ ( - \Delta + q/Q_2 )^2}{ 1 + q^2/Q^2}.
\label{eq:rllrrworst}
\end{equation}
Here 
\begin{equation} 
Q_2 = \frac{1}{2 \delta} \frac{ 1 - \sinh^2 \mu \cosh^2 \theta}{\sinh \theta \cosh \theta \cosh \mu}
\label{eq:qtilde}
\end{equation}
and
\begin{equation}
\Delta = \frac{ \sinh \mu \cosh \theta}{ 1 - \sinh^2 \mu \cosh^2 \theta}.
\label{eq:delta}
\end{equation}
Thus the reflection coefficients have a asymmetric form associated with Fano rather than 
Lorentzian lines \cite{fano}. Note that $r_L$ vanishes at $q = - \Delta Q_2$ and $r_R$ at
$q = \Delta Q_2$. Thus the reflection minima are displaced symmetrically away 
from the $q=0$ peak in the transmission. 

\section{Conclusion}
\label{sec:conclusion}

In this paper we study the role of boundary conditions in ${\cal PT}$ quantum mechanics
by consideration of a simple example. There is a well developed theoretical understanding
of the role of boundary conditions in determining whether an operator is self adjoint \cite{vonneuman, stone}, 
but the
corresponding problem in ${\cal PT}$ quantum mechanics has not previously been investigated
to our knowledge. We show that for a hermitian particle in a box the allowed boundary conditions
can be parametrized by the pair $(\lambda_1, \lambda_2)$ where $\lambda_1$ and $\lambda_2$
are real numbers. On the other hand for a ${\cal PT}$ symmetric particle in a box the allowed
boundary conditions may be parametrized as $( \ell_1 + i \ell_2, - \ell_1 + i \ell_2)$ where $\ell_1$
and $\ell_2$ are real. Thus the two sets of boundary conditions overlap for the case that $\lambda_1 = - \lambda_2$
and $\ell_2 = 0$ respectively. This case corresponds to Hamiltonians that are hermitian and separately
respect both ${\cal P}$ and ${\cal T}$. Here however we are interested in Hamiltonians that are not 
hermitian but respect the combined symmetry ${\cal PT}$. For such Hamiltonians we find that as long
as $\ell_1 > 0$, not only does the Hamiltonian commute with ${\cal PT}$, one also obtains the stronger
result that ${\cal PT}$ is unbroken. In other words one can find simultaneous eigenfunctions of the
Hamiltonian and ${\cal PT}$ and the Hamiltonian eigenvalues are necessarily real. Furthermore we
find that the Hamiltonian for a particle in a box 
with ${\cal PT}$ symmetric boundary conditions is self-adjoint under the 
${\cal PT}$ inner product. Thus the ${\cal PT}$ symmetric particle in a box fulfills all three
requirements  of ${\cal PT}$ quantum mechanics of which it constitutes a simple soluble example. 

A second development in this paper is the formulation of a variational principle for
${\cal PT}$ quantum mechanics that is the analog of the textbook Rayleigh-Ritz principle. 
The latter principle is the basis for many approximations and insights into quantum
mechanics and its generalization to ${\cal PT}$ quantum mechanics may therefore prove
of similar value. 

Finally we study classical electromagnetic analogs of the ${\cal PT}$ symmetric particle in a box.
We show that the natural photonic analog of the isolated particle in a box is a 
Fabry-Perot cavity bounded by an absorbing medium
and its conjugate gain medium. One way to experimentally realize and probe a particle in a
box is to consider a double barrier structure. In the limit of infinite barrier height this system is
an ideal particle in a box but for finite height the bound states of the particle in a box are broadened
into resonances that can be mapped out by measurements of the transmission, reflection and
absorption through the structure. In textbook quantum mechanics the transmission and reflection
resonances have a classic Breit-Wigner shape; there is no absorption. In the ${\cal PT}$ symmetric
case we find that the transmission has a Breit-Wigner shape but the reflection and absorption have
an asymmetric Fano lineshape. 

We conclude by identifying some problems that are left open. 
(a) The potential real $x^M$ tends to the ordinary particle in a box in the limit $M \rightarrow \infty$. 
The authors of ref \cite{benderbox} have studied the eigenvalues  of the ${\cal PT}$ symmetric potential 
$x^M x^{i \epsilon}$ and have derived asymptotic results in the limit $M \rightarrow \infty$ and
$\epsilon \rightarrow \infty$. It would be desirable to study the relationship between their asymptotic results
and the ${\cal PT}$ symmetric particle in a box model studied here. (b) We do not at present have
a proof of the completeness of the eigenfunctions for the non-hermitian particle in a box.
In the hermitian case the completeness can be proved by regulation and direct evaluation
of the completeness sum eq (\ref{eq:bicompleteness}) or by use of the variational principle \cite{morse}; it is guaranteed
by general theorems moreover \cite{stone}. The non-hermitian case is more subtle. In particular
the variational proof \cite{morse} does not generalize because variational principle for ${\cal PT}$ quantum
mechanics is based on the ${\cal PT}$ symmetric inner product which is indefinite. (c) Spontaneous 
${\cal PT}$ symmetry breaking is a remarkable feature of ${\cal PT}$ quantum mechanics \cite{bendereview}.
Our model may provide a simple tractable example of such a transition for
$\ell_1 < 0$.
(d) In the maximally non-hermitian case $\ell_1 = 0$ 
our model has a sequence of critical points at which it has an orthogonality catastrophe.
These critical points correspond to the circumstance that a particular eigenstate becomes
weightless (zero norm) with respect to the ${\cal PT}$ inner product, a phenomenon worthy of
further elucidation. (e) In this paper we limited attention to a non-relativistic spinless particle 
for which time-reversal symmetry is even in the sense that ${\cal T}^2 = 1$. Two natural
generalizations are to consider a particle with spin $\frac{1}{2}$ for which time reversal
symmetry is odd ${\cal T}^2 = -1$ and a particle that obeys the relativistic Dirac equation.
(f) Finally we describe a possible experimental realization of the ${\cal PT}$ symmetric particle in a
box using the Fe-doped LiNbO$_3$ system studied by ref \cite{ptexpt}. 
We envisage forming a waveguide by forming a suitable refractive index profile $n_R(x)$ 
transverse to the direction of propagation as in ref \cite{ptexpt}. 
Asymmetric optical gain can then be provided by two-wave
mixing and a suitable amplitude mask that allows the pump beam to illuminate only one
side of the waveguide. The temporal dynamics of the ${\cal PT}$-symmetric particle in a box
can be mapped out by spatial propagation of light down the waveguide, much as the temporal
dynamics of a  two level non-hermitian system was mapped out in ref \cite{ptexpt}. The experiment we envisage 
would be particularly well suited to study the ${\cal PT}$ phase transition and the orthogonality
catastrophes that occur in our system. More detailed consideration of this and other 
experimental realizations are left for future investigation.


Kate Jones-Smith is supported by the US DOE under Contract No. DE-FG02-91ER40628 and
the NSF under Grant No. PHY-0855580. Harsh Mathur
is supported by the DOE at Case Western Reserve University. 
Jason Tabachnik was supported by SURES, a summer undergraduate research 
program of Case Western Reserve University. 

\vspace{3mm}

\appendix
\label{section:app}

\section{Mapping to textbook particle in a box}
\label{app:relationship}

In this appendix we examine more closely the relationship of the maximally non-hermitian
particle in a box with boundary conditions $(i \ell_2, i \ell_2)$ to the textbook particle in a box
with boundary conditions $(0,0)$. Since the two problems are iso-spectral it is easy to find
a similarity transformation between them. Denoting the eigenfunctions of the textbook problem
\begin{equation}
\xi_n (x) = \sqrt{ \frac{2}{L} } \sin \frac{ \pi n x }{L},
\label{eq:textbookwavefns}
\end{equation}
with $n = 1, 2, 3, \ldots$, we see that the kernel
\begin{equation}
K (x, x') = \sum_{n = 1}^{\infty} \psi_n(x) \xi_n (x')
\label{eq:nonlocalkernel}
\end{equation}
maps eigenfunctions of the textbook box to the eigenfunctions of the maximally non-hermitian
problem. Here $\psi_n(x)$ is given by eq (\ref{eq:efnsptbox}). Conversely, the kernel
\begin{equation}
M(x, x') = \sum_{n = 1}^\infty \xi_n (x) \varphi_n^\ast (x')
\label{eq:inversekernel}
\end{equation}
maps eigenfunctions of the non-hermitian box back to the textbook eigenfunctions by
virtue of the biorthogonality eq (\ref{eq:biorthogonality}). The kernel $M$ is the inverse of $K$ 
in the sense that
\begin{equation}
\int d x'' \; M(x, x'') K( x'', x') = \delta ( x - x' )
\label{eq:inverse}
\end{equation}
by virtue of the completeness of the eigenfunctions of the textbook particle in a box. 
At present we lack a proof that $M$ is also the right inverse of $K$; that would be 
tantamount to a proof of bi-completeness. 

The existence of a mapping between the two problems raises the question whether they 
in fact represent the same physics in a different representation. However it can be shown that
the kernels $K$ and $M$ are non-local. Thus the perturbation of the ideal box problem that
is local in one representation will look non-local in the other. Thus the two problems may
in fact be considered physically distinct. Note also that the existence of this mapping is a
peculiarity of the maximally non-hermitian box. More general ${\cal PT}$ symmetric boundary
conditions should not be iso-spectral with any hermitian boundary conditions. 

To conclude we now prove that the kernel $K$ is non-local. If $K$ were local it would
have a delta function spike at $x = x'$ and it would vanish for $x$ and $x'$ distinct. 
Thus our task is to show that $K ( x, x' ) \neq 0$ for at least some distinct $x$ and $x'$. 
To this end it is convenient to split the expression for $K$ into a sum of two terms. The first of these terms is
\begin{eqnarray}
K_1 & = & \frac{2}{\pi \ell_2} \sum_{n = 1}^\infty \frac{1}{n} \sin \frac{ \pi n x}{L} \sin \frac{ \pi n x' }{L}
\nonumber \\
& + & i \frac{2}{L} \sum_{n=1}^\infty \sin \frac{ \pi n x'}{L} \cos \frac{\pi n x}{L}.
\label{eq:kone}
\end{eqnarray}
The convergence of these sums is a bit delicate but they may be evaluated analytically
with the result
\begin{eqnarray}
K_1 & = & \frac{i}{L} \cot \frac{ \pi(x' - x) }{L} + \frac{i}{L} \cot \frac{\pi (x'+x) }{L}
\nonumber \\
& + & \frac{1}{\pi \ell_2} \ln \left| \frac{ \sin [ \pi (x + x')/ L ]}{\sin [ \pi (x - x')/L ] } \right|.
\label{eq:koneexact}
\end{eqnarray}
Thus $K_1$ does not vanish for $x$ and $x'$ distinct. Indeed it diverges as $x \rightarrow x'$. 
We will now show that the second term, $K_2$ remains finite and therefore cannot
cancel the divergent term. Hence $K_1 + K_2$ also does not vanish at least for $x$ and $x'$
sufficiently close, showing the $K$ is indeed non-local. 

The exact expression for the second term is
\begin{eqnarray}
K_2 & =  & \frac{2}{L} \sum_{n=1}^\infty
\left[ \left| \left( \frac{ \pi \ell_2 n}{L} \right)^2 - 1 \right|^{-1/2} - 
\left( \frac{ \pi \ell_2 n}{L} \right)^{-1} \right]
\nonumber \\
& & \hspace{10mm}
\times  \sin \frac{ \pi n x}{L} \sin \frac{ \pi n x'}{L}
\nonumber \\
& + & i \frac{2}{L} \sum_{n = 1}^\infty \left[ 
\frac{ \pi \ell_2 n}{L} 
\left| \left( \frac{ \pi \ell_2 n}{L} \right)^2 - 1 \right|^{-1/2} - 1 \right] 
\nonumber \\
& & \hspace{10mm}
\times \sin \frac{ \pi n x'}{L} \cos \frac{ \pi n x}{L}.
\label{eq:ktwo}
\end{eqnarray}
Neither sum depends on the oscillatory terms for its convergence in contrast to the 
situation for $K_1$. It follows 
\begin{eqnarray}
| K_2 | & \leq & \frac{2}{L} \sum_{n=1}^\infty
\left| \left| \left( \frac{ \pi \ell_2 n}{L} \right)^2 - 1 \right|^{-1/2} - 
\left( \frac{ \pi \ell_2 n}{L} \right)^{-1} \right|
\nonumber \\
& + & \frac{2}{L} \sum_{n = 1}^\infty \left|
\frac{ \pi \ell_2 n}{L} 
\left| \left( \frac{ \pi \ell_2 n}{L} \right)^2 - 1 \right|^{-1/2} - 1 \right| 
\label{eq:ktwobound}
\end{eqnarray}
The right hand side of eq (\ref{eq:ktwobound}) is finite since the sum is highly convergent. 
It therefore provides a bound on $K_2$ that is constant (independent of $x$ and $x'$). 
This completes the demonstration that $K$ is non-local.

\end{document}